# Combined Experimental and Theoretical Studies on Iodine Capture of Zr-based Metal-Organic Frameworks: Effect of N-functionalization and Adsorption Mechanism

Jie Liang,[a,#,] Haoyi Tan,[b,#] Jiaomei Liu,[c] Huizhao Qi,[c] Xin Li,[b] Liu Wu,[c] Xiangfei Xue,[c] Guangcun Shan[b,d,*]

[a] School of Energy and Power Engineering, Beihang University, 100262, China

[b] School of Instrumentation Science and Opto-electronics Engineering, Beihang University, Beijing, 100083, China

[c] School of Space and Environment, Beihang University, 100262, China

[d] Department of Materials Science and Engineering, City University of Hong Kong, Hong Kong SAR, China

[#]The authors contributed equally to this work.

[*]Corresponding author. Email address: gcshan@buaa.edu.cn (G.C. Shan)

# Abstract

The potential leakage of nuclear waste, especially radioiodine, is a major safety concerning issue around the world. To remove radioiodine from nuclear waste efficiently, there is an urgent demand for adsorbents that possess both high stability and strong adsorption affinity for environmental remediation. Herein, two Zr-based metal-organic frameworks (Zr-MOFs) and their N-functionalized analogues have been synthesized and researched for iodine adsorption in both vapours and solutions. It was found that Zr-MOFs with N-enriched ligands (*e.g.*, pyridine and amino) exhibited the faster iodine adsorption rate and the higher iodine uptake amount (*e.g.*, reaching adsorption equilibrium within 4 hours with the removal rate of above 85% for iodine solution adsorption) than their unfunctionalized counterparts (UiO-66 and UiO-67). The critical role played by N-enriched groups in enhancing iodine adsorption has been revealed through versatile model fittings, X-ray photoelectron spectroscopy (XPS) and Raman spectroscopy characterizations, as well as density functional theory (DFT) calculations. Compared to those in amino-group, the N-atoms in pyridine-groups showed a deeper affinity towards iodine molecules. Remarkably, the N-enriched UiOs adsorbents also exhibited good recyclability, especially UiO-66-PYDC and UiO-67-$NH_2$ could maintain the removal efficiency of 89.05% and 85.49% after four adsorption-desorption recycling tests. With the strong iodine uptake affinity and outstanding regeneration performance, this work has systematically investigated the impact of N-functionalization on the enhanced performance for iodine capture by using the N-enriched UiO MOFs as promising adsorbents, providing an insightful guideline into the physical chemistry of adsorption mechanism behind the radioiodine capture.

# 1. Introduction

Given its sustainability and low-carbon emissions, nuclear energy is anticipated to play an increasingly crucial part in years to come. Concerning the nuclear power, a major safety concern is the management of nuclear wastes and the risk of potential leakage in the event of a serious malfunction or accident. The Chernobyl and Fukushima accidents witnessed the uncontrollable dispersion of hazardous radionuclides into air, soil, and ocean, which posed a dramatic threaten to ecosystem and human safety [1, 2]. Among the radionuclides, $^{129}$I and $^{131}$I isotopes are of particular concern for environmental protection [3, 4]. While $^{129}$I is challenging because of its long half-life ($t_{1/2}$=1.57×10$^7$ years), the short-lived $^{131}$I ($t_{1/2}$=8 days) is harmful owing to its high ionizing radiation (β-particles' energy = 606 keV; γ-particles' energy = 364 keV) [5]. In addition, both iodine isotopes are highly volatile and tend to bioaccumulate in the thyroid gland of humans, which may cause thyroid cancer. As a consequence, there is an urgent demand to remove radioiodine from nuclear waste stream.

Adsorption is a facial method to separate radioiodine from exhaust gas and wastewater in plant [6]. Accordingly, various absorbents have been researched and developed for efficient $I_2$ adsorption. Currently, activated carbon (AC) derived from coal or agriculture resides was widely applied as matrix in the nuclear power plant. However, it was noted that the AC had to be impregnated with metallic compounds (*e.g.*, KI, PI$_2$) or organic amine (*e.g.*, diethanol amine and triethylene diamine) before being used, which complicated the adsorption process and made the adsorbent regeneration difficult [7-9]. Ag-exchanged zeolites were also promising in $I_2$ capture due to the formation of AgI precipitate ($K_{sp}$ = 8.52 × 10$^{-17}$) [10, 11]. Yet, the $I_2$ uptake capacity was still lower than 200 mg/g, and the randomly distributed Ag$^+$ ions limited the $I_2$ uptake rate. Recently, metal-organic frameworks (MOFs) were demonstrated to be highly promising candidates for $I_2$ capture because of high surface area, uniform pore distribution, and versatile functionalization [12, 13]. The first utilized MOF for $I_2$ capture was ZIF-8. Benefitting from a similar pore size with the diameter of $I_2$ molecules (~5 Å), ZIF-8 exhibited a high $I_2$ uptake of ~125 wt%, of which 25 wt% $I_2$ was absorbed on the outer surface, and the left $I_2$ was trapped in the inside sodalite cages [14, 15]. It was founded that the iodine adsorbing energy on ZIF-8 was 4 times stronger than that on AC. Except for pore size control, the π-electron conjugated organic linkers also favoured $I_2$ capture. For instance, Cu-BTC showed an $I_2$ uptake of ~175 wt% in a mixed water-iodine gas stream, benefitting from the interaction between $I_2$ and benzene tricarboxylate [16]. A major breakthrough reported for MOFs was the functionalization of organic ligands by basic groups (*e.g.*, -$NH_2$, -OH), which could donate electrons and therefore boosted the affinity for electron-deficient $I_2$ by forming a charge transfer complex [17]. Taking advantage of this strategy, MIL-53-$NH_2$ removed 60% $I_2$ from the iodine/cyclohexane solution after 48 h, 12 folds of that for MIL-53 (5%). Although the bivalent metal-based MOFs (*e.g.*, $Zn^{2+}$ and $Cu^{2+}$) demonstrated excellent iodine adsorption property, their low stability limited their application in real nuclear waste disposal environment, *e.g.*, in hot vapor or in acid aqueous solution [18]. In this context, MOFs composed of trivalent metals or higher, more resistant to harsh conditions, could be more favorable for $I_2$ capture from nuclear waste.

Generally, Zr-MOFs demonstrated a high stability owing to the strong metal-ligand bonds [19-21]. For instance, the UiO series (*e.g.*, UiO-66) of Zr-MOFs not only kept their structural integrity up to 500 ºC, they were also stable in aqueous and acidic solutions (pH=1) [22]. Apart from the structure flexibility of MOFs, Zr- MOFs could also be used for the geological sequestration of radioiodine over long periods of time [23]. However, despite these advantages, the investigations on the application of Zr-MOFs in radioiodine capture were limited; and the interactions between guest iodine and different Zr-MOFs hosts were also not clear [18, 24, 25]. The aim of this work is to expand the range of available Zr-MOFs adsorbents for radioiodine removal application and reveal the adsorption mechanism. Firstly, two series of Zr-MOFs, including the landmark UiO-66, UiO-67 and their functionalized analogies (*e.g.*, UiO-66-$NH_2$, UiO-66-PYDC, UiO-67-$NH_2$, UiO-67-BPYDC) were synthesized and tested for gaseous and soluble iodine adsorption under different conditions.

Then, the recyclability of N-functionalized UiOs in soluble iodine solutions was evaluated. Moreover, to study the interactions between guest iodine and N-functionalized UiOs, detailed spectroscopy characterizations as well as DFT calculations were conducted for revealing the adsorption mechanism.

## 2. Methods

### 2.1. Synthesis of Zr-MOFs

The synthetic methods for different Zr-MOFs were established and/or modified according to previous published literatures [26].

For UiO-66, UiO-66-NH$_2$ and UiO-67: 150 mg ZrCl$_4$ (0.643 mmol) and the corresponding organic linker - 106 mg terephthalic acid (BDC, 0.638 mmol), 116 mg 2-aminoterephthalic acid (BDC-NH$_2$, 0.640 mmol) or 155 mg biphenyl-4,4'-dicarboxylic acid (BPDC, 0.635 mmol), were mixed in 10 mL dimethyl formamide (DMF) under ultrasonic. The above reactants were then loaded into a 25 mL Teflon-lined autoclave and heated at 120 °C for a period of 24 h. After crystallization, the as-synthesized solid product was separated by centrifugation, washed three times with 10 mL DMF, and then heated in 15 mL DMF at 60 °C for 12 h to remove free ligands. After cooling down, the Zr-MOFs were centrifuged, washed three times with 10 mL acetone, and heated in 15 mL acetone at 60 °C for 72 h to get rid of the trapped DMF solvent. Finally, the pure Zr-MOFs were obtained after drying at 80 °C under vacuum for 24 h. Corresponding to the ligands in turn, the obtained Zr-MOFs adsorbents were UiO-66, UiO-66-NH$_2$ and UiO-67, respectively.

For UiO-66-PYDC: 233 mg ZrCl$_4$ (1.000 mmol), 168 mg pyridine-dicarboxylic acid (PYDC, 1.005 mmol) and 4 mL acetic acid (99.7% wt%) were dispersed and mixed in 6 mL aqueous solution. The above reactants were loaded into an autoclave and heated at 100 °C for 24 h. Then, the as-synthesized UiO-66-PYDC powder was solvent-exchanged three time by 10 mL DMF and 10 mL acetone, respectively. And the final UiO-66-PYDC adsorbent was collected after heated in vacuum at 80 °C for 24 h.

For UiO-67-NH$_2$: 167.5 mg ZrCl$_4$ (0.719 mmol) and 1.25 mL hydrochloric acid (HCl) were dissolved in 22.5 mL DMF under ultrasonic. 230 mg 2-amino-biphenyl-4,4'-dicarboxylic acid (BPDC-NH$_2$, 0.894 mmol) was dissolved in 15 mL DMF. The above DMF solution was mixed and heated at 80 °C for 24 h. After filtrating, the white powder was sequentially treated three times by 10 mL DMF and 10 mL acetone, respectively, and finally activated at 80 °C in vacuum for 24 h to yield the UiO-67-NH$_2$ adsorbent.

For UiO-67-BPYDC: 150 mg ZrCl$_4$ (0.643 mmol) and 157 mg 2,2'-bipyridine-5,5'-dicarboxylic acid (BPYDC, 0.643 mmol) were dissolved in DMF (10 mL)/HCl (0.5 mL) under ultrasonic. The reactants were heated at 60 °C for 24 h. After crystallization, the resultant solid product was washed three times by 10 mL DMF and 10 mL acetone. Finally, UiO-67-BPYDC was obtained after heated under vacuum at 80 °C for 24 h.

**2.2. Characterizations**

Powder X-ray diffraction (PXRD) patterns were characterized from 5° to 30° (2$\theta$) at a scan rate of 5°/min on a Bruker D8 Advance X-ray diffractometer (Cu-K$\alpha$ radiation source, $\lambda$=1.54184Å, 40 kV, 60 mA). The morphology of Zr-MOFs was evaluated on a ZEISS Germini sigma-500 scanning electron microscope (SEM) with an accelerating voltage of 3 kV. The N$_2$ sorption experiments were applied using a micromeritics ASAP 2020 plus gas adsorption analyzer at 77 K. Before adsorption, the samples were degassed at 150 °C for 6 hours. The pore size distribution was obtained according to non-local density functional theory (NLDFT) method. UV-vis absorption spectra were collected using an ultraviolet spectrophotometer (Hitachi U-3900 H) with the wavelength range from 200 to 700 nm. Raman spectra were collected by applying a Renishaw inVia spectrometer with the excitation wavelength set as 532 nm. The XPS spectra were characterized on a Thermo Scientific Escalab 250Xi electron spectrometer with monochromatic Al K$\alpha$ radiation (C 1s at 284.8 eV).

**2.3. Adsorption Experiments**

For gaseous iodine adsorption experiments, 30 mg Zr-MOF was placed in 8 mL vial, which was then weighted and transferred to a 100 mL wide-mouth bottle containing 300 mg iodine solid. The whole bottle was airtight and heated at 80 °C. The schematic diagram of above experiments was plotted in Fig. S1. After a desired time, the bottle was opened up, the inside vial was taken out and treated at 80 °C for another 2 min to take out the free iodine vapor. After cooling down, the vial containing Zr-MOF adsorbents was weighted till no weight change. Subsequently, the iodine adsorption capacity Q (g/g) for each Zr-MOF could be calculated as below:

$$Q = \frac{m_2 - m_1}{m_1} \quad (1)$$

in which $m_1$ and $m_2$ represent the initial and final mass of vial (with Zr-MOFs), respectively.

For $I_2$ solution adsorption experiments, 45 mg Zr-MOF adsorbent was put in a vial containing 10 mL $I_2$/cyclohexane solution. The sealed vial was then shaken at 25 °C at a speed of 250 rpm. The adsorption kinetics were studied in 0.5 mmol/L iodine solution with adsorption time ranging from 2 h to 24 h. The adsorption isotherms were collected by changing the concentration of $I_2$ solution from 0.5 to 25 mmol/L with the adsorption time fixed at 24 h for an equilibrium guarantee. After adsorption, the absorbent was separated after centrifugation, and the $I_2$ concentration in supernatant was measured on the ultraviolet spectrophotometer with the wavelength of 522 nm. In order to guarantee the reliability, each experiment (whether for gaseous iodine adsorption or $I_2$ solution adsorption) was conducted a minimum of two times.

The equilibrium adsorption amount ($q_e$, mg/g) and adsorption efficiency ($R$, %) were obtained as below:

$$q_e = \frac{(c_0 - c_e) \times V}{m} \quad (2)$$

$$R = \frac{c_0 - c_e}{c_0} \times 100\% \quad (3)$$

in which $c_0$ (mg/L), $c_e$ (mg/L) and $V$ (mL) respectively represent the initial concentration, equilibrium concentration and the volume of $I_2$ solution; and $m$ (mg) denotes the mass of fresh Zr-MOF adsorbent.

The pseudo-first-order (**Eq. 4**) and pseudo-second-order models (**Eq. 5**) were employed according to the following equations: (The non-linear forms of adsorption kinetics models were shown in supplementary material as **Eq. S1 and Eq. S2**)

$$\log(q_e - q_t) = \log q_e - \frac{k_1}{2.303}t \quad (4)$$

$$\frac{t}{q_t} = \frac{1}{k_2 q_e^2} + \frac{t}{q_e} \quad (5)$$

in which $q_e$ (mg/g) and $q_t$ (mg/g) represent the uptake amount at equilibrium stage and time $t$ (h), respectively; $k_1$ and $k_2$ denote the pseudo-first-order and pseudo-second-order adsorption rate constants.

The adsorption isotherm fitting was respectively based on Langmuir (**Eq. 6**) and Freundlich (**Eq. 7**) models: (The non-linear forms of adsorption isotherm models were shown in supplementary material as **Eq. S3 and Eq. S4**)

$$\frac{C_e}{q_e} = \frac{1}{K_L q_{max}} + \frac{C_e}{q_{max}} \quad (6)$$

in which $c_e$ (mg/L) and $q_e$ (mg/g) respectively represent the concentration and uptake amount at equilibrium stage; $q_{max}$ (mg/g) stands for the maximum uptake amount; and $K_L$ (L/mg) denotes the Langmuir constant.

$$\log q_e = \log k_F + (n)\log C_e \quad (7)$$

in which $k_F$ (mg$^{1-n}$ L$^n$ g$^{-1}$) and $n$ are the Freundlich constant and heterogeneity factor, respectively.

The external mass transfer diffusion model (**Eq. 8**) was used to fit the experimental data as below equation [29, 30]:

$$\left[\frac{d(c/c_0)}{dt}\right]_{t\to 0} = -\beta_L S \quad (8)$$

in which $c_0$ (mg/L) and $c$ (mg/L) respectively stand for the concentration of I$_2$ solution at initial stage and time $t$; $S$ (cm$^2$/cm$^3$) is the volumetric surface of adsorbent, and $\beta_L$ (cm/s) denotes the external mass transfer diffusion coefficient.

The intraparticle mass transfer models developed by Weber & Morris and McKay & Poots (**Eq. 9**) and Urano & Tachikawa (**Eq. 10**) were fitted according to the following equations, separately [31-33]:

$$q = Kt^{0.5} + C \quad (9)$$

$$f\left(\frac{q}{q_m}\right) = -\left[\log_{10}\left(1 - \left(\frac{q}{q_m}\right)^2\right)\right] = \frac{4\pi^2 D_i t}{2.3 d_p^2} \quad (10)$$

in which $q$ (mmol/g) and $q_m$ (mmol/L) respectively represent the uptake amount at time $t$ (min) and at equilibrium stage; $C$ is the intercept; $d_p$ (cm) denotes the particle diameter of Zr-MOFs; K (mmol/g/s) and $D_i$ (cm$^2$/s) represent the diffusion coefficient inside the adsorbent. Significantly, the linearisation of McKay & Poots model was fitted based on the initial stage of adsorption within the first 240 min.

To investigate the reusability of Zr-MOFs, 45 mg adsorbent loaded with iodine was first washed three times by 10 mL ethanol and then stirred in ethanol for 36 h. After centrifugation, the regenerated adsorbent was activated under vacuum at 80°C for 24 h before the next adsorption cycle.

### 2.4. DFT calculations

For revealing the adsorption mechanism of I$_2$ on Zr-MOFs, DFT calculations were carried out using the Vienna *ab initio* simulation package (VASP). The interaction energy between iodine molecules and different linkers was hence calculated. The Perdew-Burke-Ernzerhof (PBE) functional in the generalized gradient approximation (GGA) was performed to describe the electron-ion interactions [32-35]. To consider the van der Waals interactions during the adsorption process, DFT-D3 corrections were also included [36]. The cut-off energy for the plane wave basis was set as 450 eV and only the Gamma point was used to sample the first Brillouin zone because of the large size of the unit cells. The accuracy of electronic self-consistency was set to 10$^{-6}$ eV between the two electronic steps. The structures were fully relaxed by converging the residual forces between atoms lower than 0.001 eV/Å. Additionally, a vacuum layer thicker than 15 Å was added to prevent the disturbances during the calculations. Finally, Gaussian smearing with σ = 0.1 eV was used to help electronic convergence better.

In addition, the adsorption energy of I$_2$ was estimated as below equation:

$$\Delta E = E_{coordination} - E_{I_2} - E_{linker}$$

in which $E_{coordination}$ stands for the optimized energy of the coordinated structures after fully structure relaxation, $E_{I_2}$ and $E_{linker}$ are energy of the individual I$_2$ and linker, respectively [25].

## 3. Results and discussions

### 3.1. Characterization of Zr-MOFs

Two landmark Zr-MOFs (UiO-66, UiO-67) and their N-functionalized homologs (UiO-66-NH$_2$, UiO-66-PYDC, UiO-67-NH$_2$, UiO-67-BPYDC) were synthesized by using BDC, BPDC, BDC-NH$_2$, PYDC, BPDC-NH$_2$ and BPYDC as the organic ligands (**Fig. 1**a and c), respectively. All PXRD patterns of above Zr-MOFs exhibited the characteristic peaks of reported UiO-66 or UiO-67 [26], in which the absence of additional peaks confirmed their phase purity (**Fig. 1**b). In addition, the almost identical peak positions for the UiO-66 or UiO-67 series suggested that the utilization of N-enriched ligands barely changed the framework of Zr-MOFs. Taking the UiO-66 for example, the structure was shown in **Fig. 1**a. SEM images (Fig. S1) showed that all Zr-MOFs were polycrystalline particles (~100 nm), among which were sparsely decorated by several discrete octahedral crystals (200~500 nm).

The porosity properties of Zr-MOFs were demonstrated by N$_2$ sorption isotherms (**Fig. 1**d). All Zr-MOFs exhibited a typical type I isotherm, as evidenced from the steeply increased adsorption amount under the relatively lower pressure ($0.001 < P/P_0 < 0.01$). The warped-up tail in the higher pressure ($0.85 < P/P_0 < 1.0$) of UiO-66-PYDC and UiO-67-BPYDC was ascribed to inter-crystalline pores and consistent with their SEM image (Fig. S1). The six Zr-MOFs all possessed two kinds of pores in the framework, as demonstrated by the green and yellow spheres in **Fig. 1**a. The specific pore size distribution was plotted in Fig. S2, all of which were dominated by micropores. Table S1 summarized the texture properties of Zr-MOFs. As shown, the surface area and pore volume for UiO-66 were 1308 m$^2$/g and 0.441 m$^3$/g, while those for UiO-67 were increased to 1617 m$^2$/g and 0.651 m$^3$/g, respectively, owing to the extended linker length in UiO-67. The substitution of N-enriched linker decreased both values. For instance, the surface area and pore volume for UiO-66-NH$_2$ and UiO-66-PYDC were decreased to 742, 673 m$^2$/g and 0.299, 0.290 m$^3$/g, while those for UiO-67-NH$_2$ and UiO-67-BPYDC were reduced to 863, 755 m$^2$/g and 0.421, 0.340 m$^3$/g.

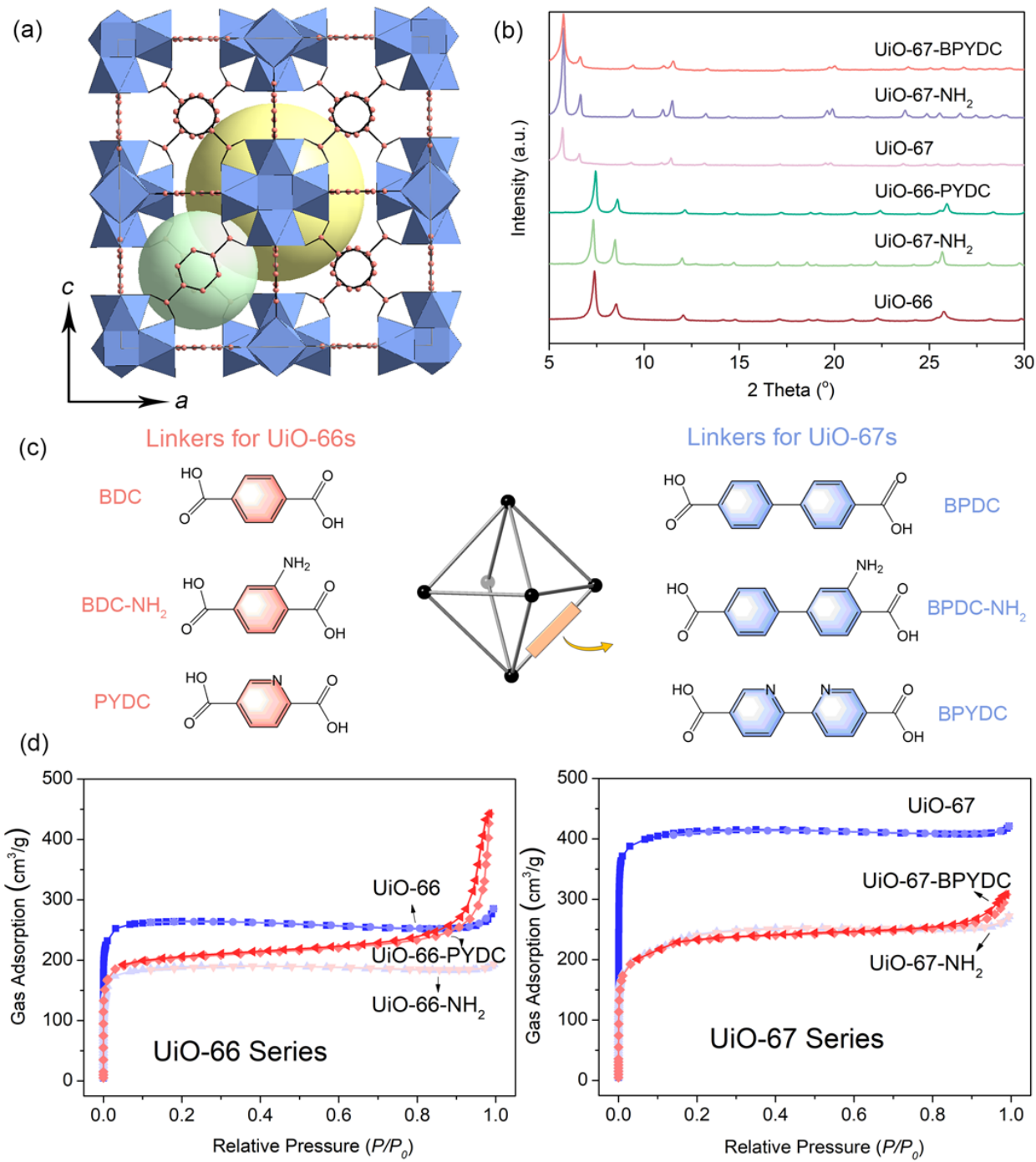

**Fig. 1.** Structural characteristics of UiOs. (a) Structure of UiO-66, (b) PXRD of as-synthesized UiOs, (c) the organic linkers of UiOs and (d) N$_2$ sorption isotherms of UiOs.

## 3.2. Gaseous Iodine Adsorption

The adsorption capability of Zr-MOFs towards gaseous iodine was first investigated. **Fig. 2**a and b indicated that both the unmodified UiO-66 and UiO-67 adsorbents had a gently increased iodine uptake curve. That is, the uptake amount of gaseous iodine was slowly increased to 0.92 and 0.81 mg/g (at 24 h) for UiO-66 and UiO-67, respectively. And their corresponding maximum iodine uptake was 1.38 and 1.29 mg/g at 48 h. The N-enriched functional groups accelerated the iodine adsorption rates and boosted the adsorption amount. Specially, the iodine uptake amount of UiO-66-NH$_2$ and UiO-66-PYDC achieved 1.80 and 2.11 mg/g after 24 h adsorption (**Fig. 2**a and b), respectively, which was 2.0 and 2.3 times of that for UiO-66 (0.92 mg/g). With the extension of time, the iodine uptake was continuously increased to 2.01 and 2.21 mg/g at 48 h, respectively. It should be noted that the gaseous iodine uptake amount was not positively related to the surface area of Zr-MOFs. That is, although the N-enriched UiO-66 exhibited a lower surface area compared to unmodified UiO-66 (Table S1), their iodine uptake capability was greatly enhanced owing to the introduction of N-enriched groups (-NH$_2$, -pyridine), which formed charge transfer complex with gaseous iodine [39]. Similarly, the UiO-67-NH$_2$ and UiO-67-BPYDC built by N-enriched ligands also respectively boosted the gaseous iodine uptake to 1.76 and 2.04 mg/g at 24 h, and further to 1.91 and 2.17 mg/g at 48 h, much higher than that of UiO-67 (1.11 mg/g at 24 h and 1.29 mg/g at 48h) under the same condition (**Fig. 2**b). The inferior iodine uptake capacity for the UiO-67 series than that for UiO-66 series confirmed again that the iodine uptake capability was not directly related to the surface area of adsorbents.

All the kinetic curves were well fitted according to the pseudo-second-order model ($R^2$ > 0.97) (**Fig. 2**c and d, Fig. S4). To obtain more accurate kinetic model parameters, non-linear adsorption kinetics models have also been considered for comparison (Fig. S5) [40-43]. The related kinetic parameters were summarized in Table S2 and Table S3.

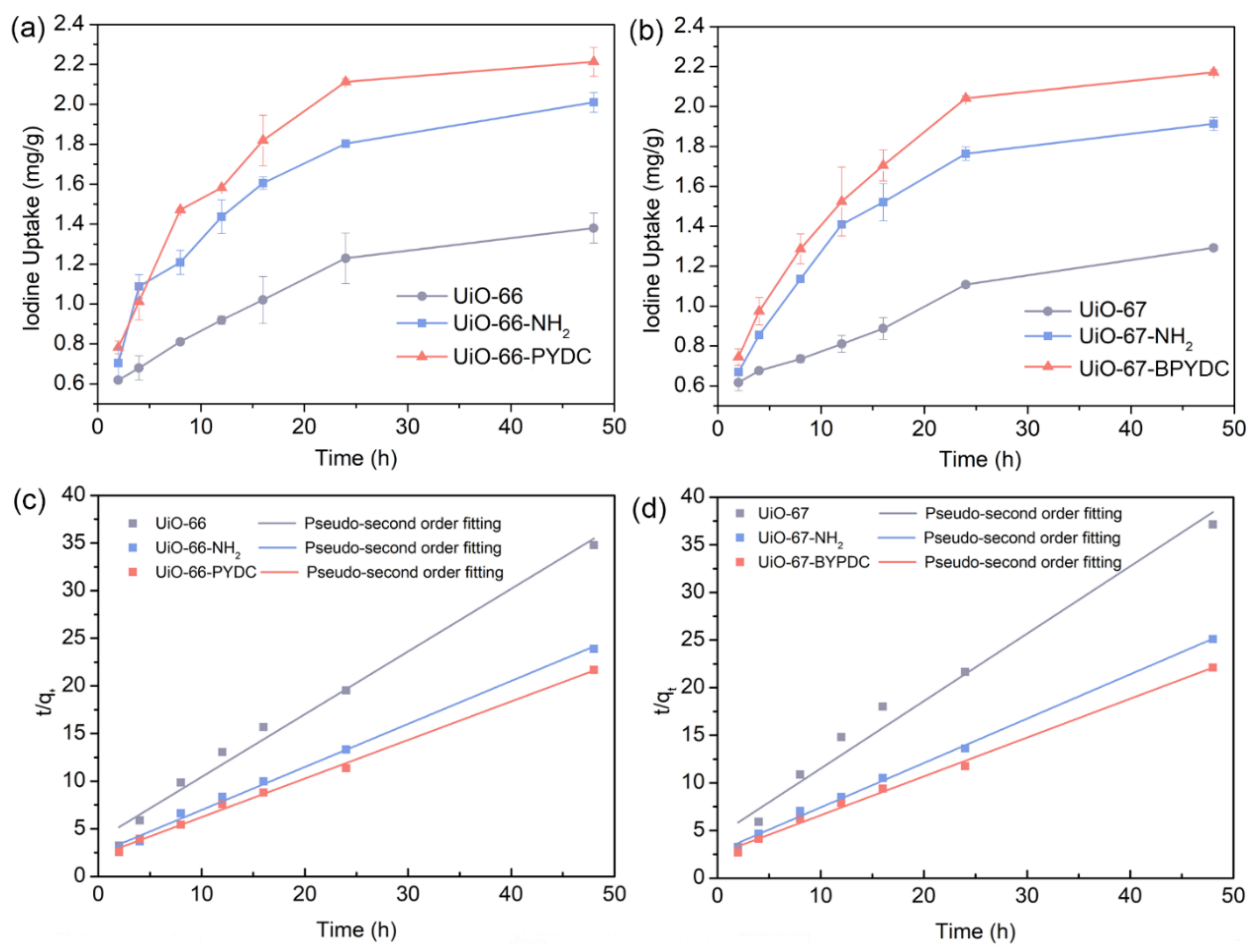

**Fig. 2.** The adsorption kinetics of Zr-MOFs towards gaseous iodine.

## 3.3. Adsorption Kinetics in Iodine Solution

Adsorption kinetics of Zr-MOFs were also conducted in 0.5 mmol L$^{-1}$ I$_2$/cyclohexane solution. The I$_2$ removal efficiency and uptake amount in function of time were depicted in **Fig. 3**. As shown, UiO-66 gradually removed 45% of I$_2$ in the initial 8 hours, after which the removal efficiency curve barely changed. UiO-67 displayed a similar I$_2$ uptake variation trend, while its I$_2$ removal efficiency (38%) at 8 hours was 7% lower than that of UiO-66. The maximum I$_2$ removal efficiency obtained by UiO-67 (46% at 24 hours) was similar to that of UiO-66 (47% at 24 hours), indicating once again that the ligand length had a minor effect on I$_2$ adsorption capacity, which was just consistent with the previous gaseous iodine adsorption experiments.

The decoration of organic ligands by N-enriched groups was conducive to I$_2$ uptake. As shown in **Fig. 3**, the N-enriched UiOs demonstrated the faster iodine adsorption rate and the higher iodine uptake amount than the unfunctionalized UiOs. In particular, both UiO-66-NH$_2$ and UiO-66-PYDC rapidly remove 80~90% of I$_2$ at the initial stage (0~2 h), which was 3.6 times greater than UiO-66 (Fig. 3a). Their I$_2$ capture efficiency was also boosted by 1.88~1.94 times (at 24 h) as compared to UiO-66. A similar phenomenon occurred in the UiO-67 series (Fig. 3b). That is, the iodine removal rate for N-enriched UiO-67 analogues was greatly increased, and more than 90% of the equilibrium uptake amount was reached for UiO-67-NH$_2$ and UiO-67-BPYDC within the first 8 hours. These promising results might be attributed to the fact that the N-enriched groups were good electron donors, which tended to form strong charge-transfer complexes with electron-deficient iodine and thereby enhanced the iodine removal from solution [15]. In both series of UiOs, it was observed that a more obvious I$_2$ uptake enhancement was observed in the pyridine-functionalized UiOs than in the NH$_2$-functionalized UiOs. That was, the UiO-66-PYDC and UiO-67-BPYDC both exhibited the faster iodine adsorption rate (92.56% and 91.23% at 2 h) than UiO-66-NH$_2$ and UiO-67-NH$_2$ (89.22% and 86.96% at 2h), respectively. This might be ascribed to a stronger charge transfer between the pyridine-functionalized ligands and iodine molecules. The slightly superior I$_2$ removal efficiency for UiO-67-BPYDC (92.74% at 24 h) relative to UiO-66-PYDC (92.12% at 24 h) could be explained by the more concentrated electron (N) density in the BPYDC ligand, which also provided more reaction sites for iodine adsorption.

The kinetic iodine adsorption data in I$_2$/cyclohexane solution was further studied by fitting based on pseudo-first and pseudo-second order models. Related parameters including correlation coefficients ($R^2$) were listed in Table S4. Based on the evaluation of $R^2$ values and fitting results (Fig. 3c and d, Fig. S6), it was evidenced that the pseudo-second order model provided a better description of the adsorption kinetics for all UiOs adsorbents. The non-linear kinetic models and parameters were shown in Fig. S7 and Table S5, respectively. The larger kinetic rates and uptake amount at equilibrium stage ($q_e$) for N-enriched UiOs relative to unfunctionalized UiOs further suggested that the N-enriched groups enhanced the interaction between iodine molecules and UiOs [44].

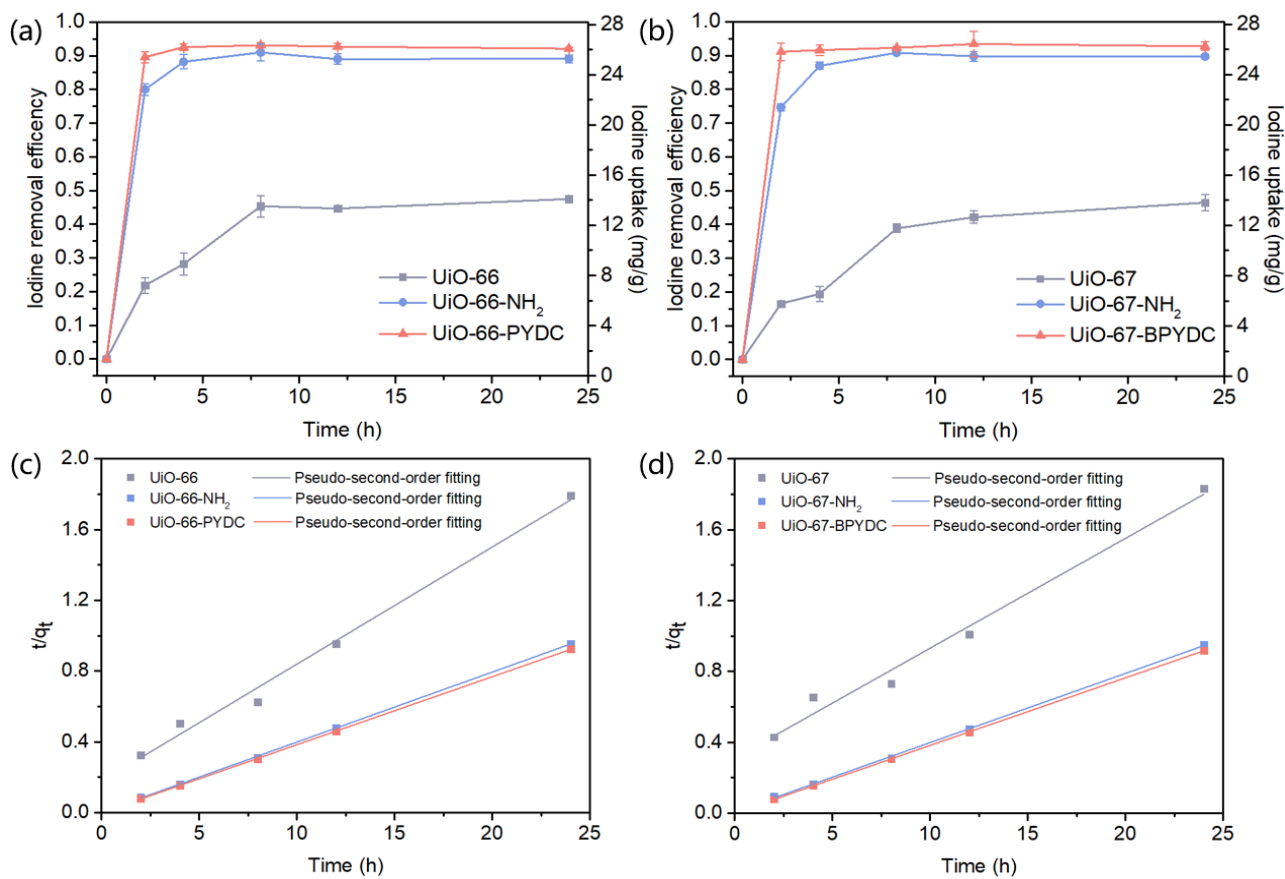

**Fig. 3.** Adsorption kinetics of (a, c) UiO-66s and (b, d) UiO-67s in iodine solution ($C_{initial}$ = 0.5 mmol/L).

### 3.4. Iodine Solution Adsorption Isotherms

The adsorption isotherms of N-enriched UiOs with different initial iodine concentrations were further collected, aiming to evaluate their loading capacities of iodine. As the $I_2$ concentration increased from 0.5 to 25 mmol/L, the equilibrium uptake amount ($q_e$) of UiO-66-NH$_2$ and UiO-67-BPYDC was gradually increased from ~25 mg/g to 273 and 342 mg/g, respectively, whereas that for UiO-66-PYDC and UiO-67-NH$_2$ was jump to above 850 mg/g (**Fig. 4**a). This indicated that although the N-enriched UiOs exhibited a similar adsorption behavior at the low $I_2$ concentration solution, the latter two adsorbents (UiO-66-PYDC, UiO-67-NH$_2$) were more advantageous in the high concentration. The Langmuir and Freundlich models were employed to fit the adsorption isotherms (**Fig. 4**b and c) [45-47], and the related parameters were listed in Table S6. Specially, the adsorption curves of UiO-66-NH$_2$, UiO-66-PYDC, UiO-67-NH$_2$ and UiO-67-BPYDC were better fitted by the Langmuir model ($R^2$ > 0.95), which suggested the iodine adsorption occurred as a monolayer process on adsorbents. The maximum adsorption capacity ($q_{max}$) was increased in the order of UiO-66-NH$_2$ (276.73 mg/g) < UiO-67-BPYDC (362.21 mg/g) < UiO-66-PYDC (919.50 mg/g) < UiO-67-NH$_2$ (1143.98 mg/g). Besides, the non-linear adsorption isotherms and related parameters were shown for comparison in Fig. S8 and Table S7 [40, 48]. The photographs of iodine/cyclohexane solution ($C_{initial}$ = 5 mmol/L) after 24 hours adsorption by all the UiOs adsorbents were displayed in **Fig. 4**a, right. It was obvious that the color of N-functionalized UiOs was much shallow than that of non-functionalized counterparts, which was direct evidence of their higher adsorption capacity. Considering the fast adsorption rates, high adsorption capacities, as well as simple synthesis methods, the N-enriched UiOs (especially UiO-66-PYDC and UiO-67-NH$_2$) could work as promising adsorbents for the treatment of radioiodine-contaminated wastewater. Furthermore, the commercial activated carbon was also tested for iodine adsorption in iodine/cyclohexane solution to verify the superiority of N-enriched UiOs as iodine adsorbents (Fig. S9). And the iodine adsorption performance in the co-presence of competing species was evaluated in Fig. S10 [49, 50].

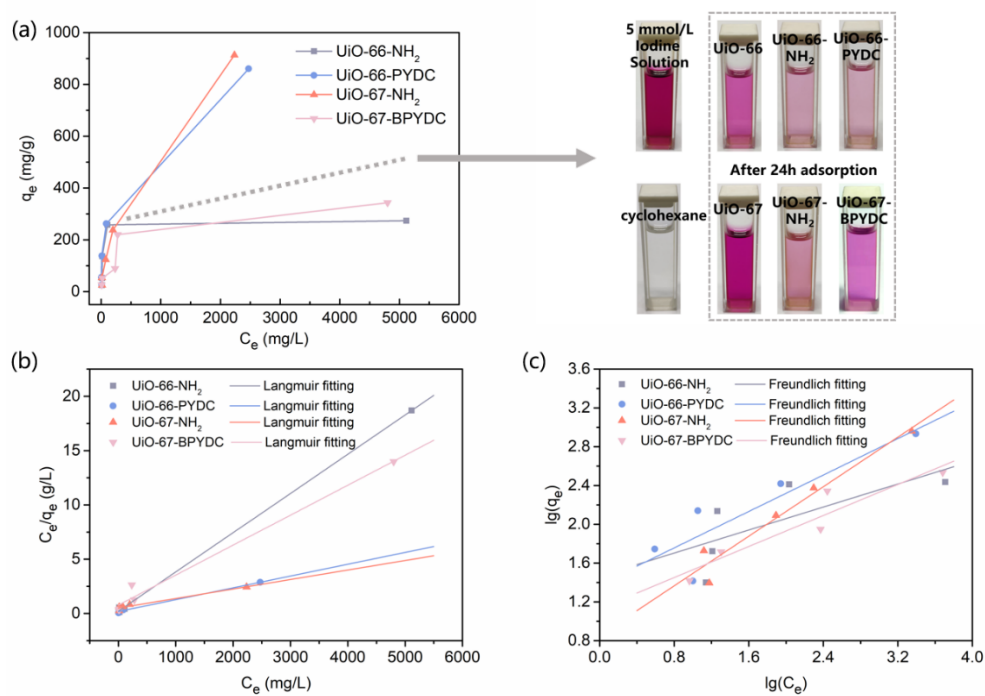

**Fig. 4.** (a) Adsorption isotherms of UiOs for the adsorption in iodine solution ($C_{initial}$ = 0.5 ~25 mmol/L), and the photographs of solution (5 mmol/L) after adsorption for 24h. (b) The fitting plots based on the Langmuir model and (c) the fitting plots based on the Freundlich model.

The mass transfer diffusion processes were also analysed to understand the adsorption behaviour of UiOs for aiding engineering applications. The iodine adsorption processes could be divided into three consecutive steps [51-53]: (1) External mass transfer diffusion process: iodine molecules transferring across the boundary liquid film around adsorbents; (2) Intraparticle mass transfer diffusion process: iodine molecules transferring from the surface into the pores of porous adsorbents; (3) Actual adsorption of iodine molecules on the active sites within or on the surface of adsorbents. The last step, often accompanied with electrostatic interactions, was much faster than the former two steps and therefore could be negligible; in other words, it was the external and intraparticle mass transfer diffusion processes that actually determined the iodine adsorption rate. To clarify the specific mass transfer diffusion processes during iodine adsorption, the experimental results were fitted using several mass transfer diffusion models, as illustrated in Fig. S11. The calculated mass transfer diffusion coefficients were summarized in Table S8. External mass transfer diffusion model was applied in Fig. S11a and b. The external mass transfer diffusion coefficient $\beta_L$ was estimated by analyzing the slope at the initial stage of plot. Based on the intraparticle mass transfer model of Weber and Morris shown in Fig. S11c and d, there were two linear portions, which was consistent with previous literatures [30, 54, 55]. It was indicated that there were two distinct adsorption stages, consisting of surface adsorption and intraparticle diffusion processes. For N-enriched UiOs, above 0.08 mmol/g iodine was adsorbed within 15 $min^{1/2}$, owing to the abundant active sites on their surface. Thereby, it could be concluded that iodine adsorption rate was limited by both external mass transfer and intraparticle diffusion processes. As shown in Fig. S11e and f based on the intraparticle mass transfer model of Urano & Tachikawa, the intraparticle mass transfer diffusion coefficient $D_i$ could be deduced from the slope at the initial stage of adsorption within 240 min. For UiO-66 series, $D_i$ was increased in the order of UiO-66 ($0.76\times10^{-6}$ $cm^2/s$) < UiO-66-$NH_2$ ($4.94\times10^{-6}$ $cm^2/s$) < UiO-66-PYDC ($7.81\times10^{-6}$ $cm^2/s$). The UiO-67 series showed a similar tendency: UiO-67 ($0.43\times10^{-6}$ $cm^2/s$) < UiO-67-$NH_2$ ($4.37\times10^{-6}$ $cm^2/s$) < UiO-67-BPYDC ($5.76\times10^{-6}$ $cm^2/s$). It was concluded that the N-containing linkers endowed Zr-MOFs a faster intraparticle mass transfer rate, and a more obvious rate increasing was observed in pyridine-functionalized UiOs than in NH$_2$-functionalized UiOs.

**3.5. Mechanism of Iodine Adsorption on N-enriched UiOs**

The outstanding performance of N-enriched UiOs motivated us to investigate the adsorption mechanism of MOF hosts towards iodine molecules. As shown in Fig. 5, Raman spectroscopy was applied to determine the characteristic of adsorbed iodine species. The peak located at 112 cm$^{-1}$ in the Raman spectra was ascribed to the formation of I$_3^-$ species. This confirmed that electron transferred between guest iodine molecules and UiOs during adsorption. Moreover, the relatively higher I$_3^-$ peak intensity for N-enriched UiOs@I$_2$ than that for unfunctionalized UiOs@I$_2$ suggested that the reactive N-atoms increased the conjugated π-electron density in linkers and hence enhanced the electron transfer between iodine and MOF hosts. For the N-enriched UiOs@I$_2$, it was observed that an extra peak of 164 cm$^{-1}$ appeared in the Raman spectra, whose maximum intensity was observed for UiO-66-PYDC@I$_2$. This peak was ascribed to the formation of another polyiodide [I$_3^-$·I$_2$] (I$_5^-$) species that was absent in the unfunctionalized UiOs@I$_2$ [56, 57].

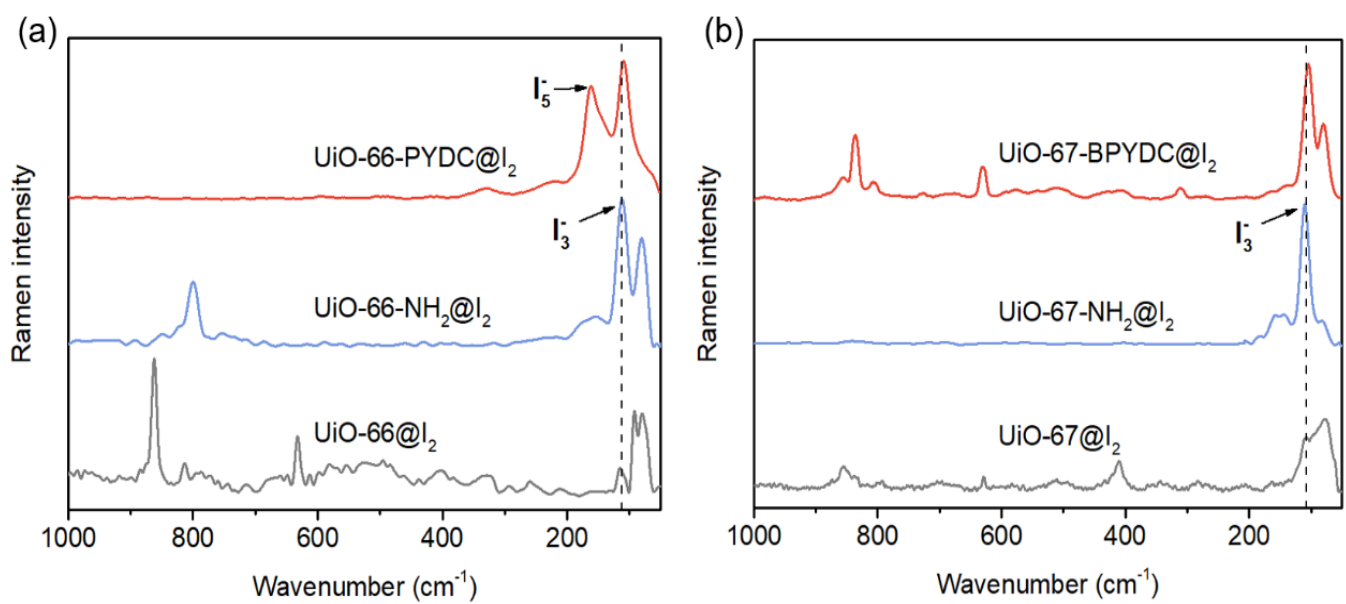

**Fig. 5.** Raman spectra of (a) UiO-66 series and (b) UiO-67 series after iodine adsorption.

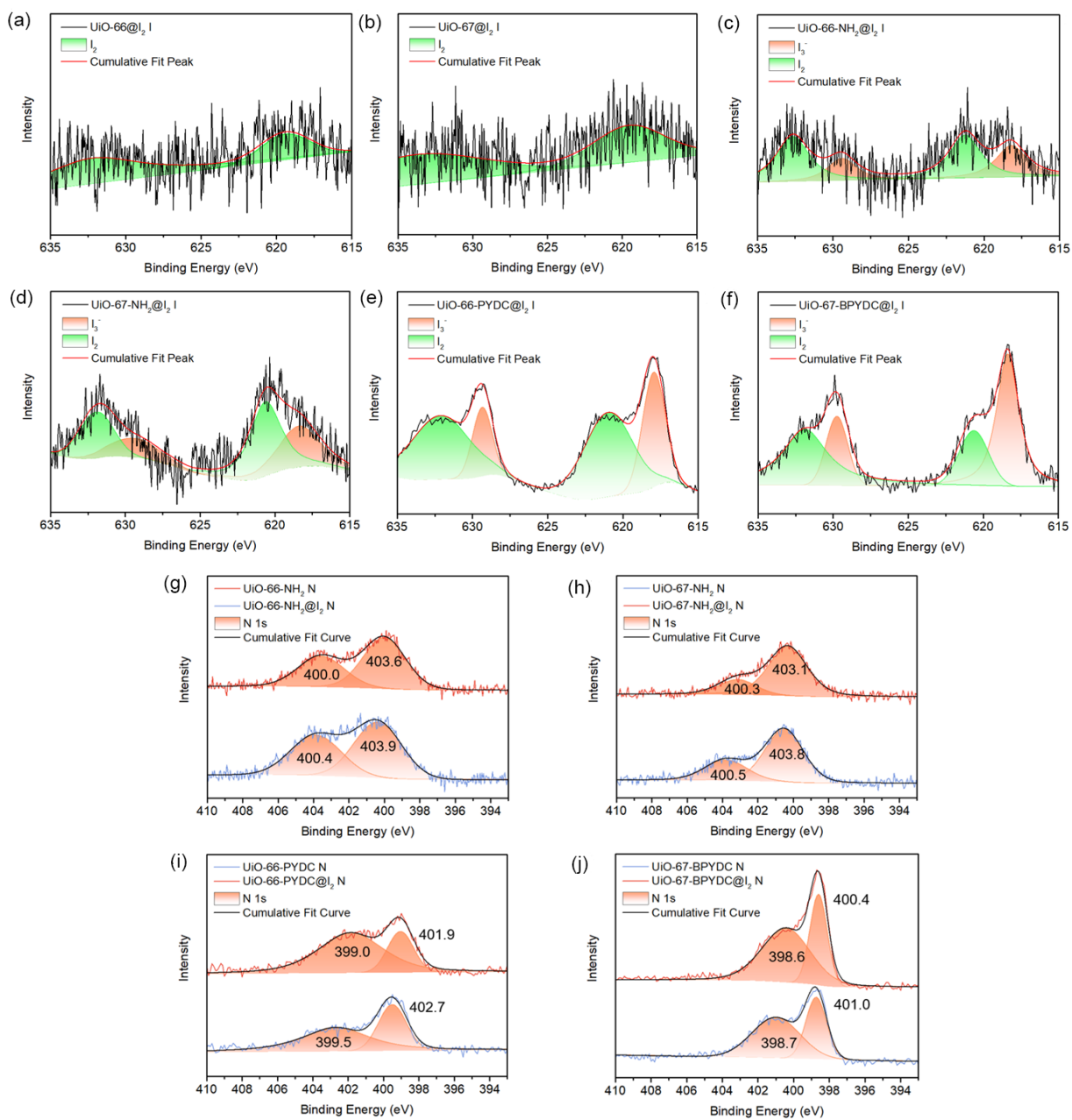

**Fig. 6.** XPS spectra of (a-f) I 3d and (g-j) N 1s for UiO-66 series and UiO-67 series.

XPS spectra was collected to clarify the bonding nature between I$_2$ and UiOs adsorbents. As shown, all the iodine loaded adsorbents exhibited two peaks positioned at 616~623 eV and 628~634 eV (Fig. 6a-f), ascribing to I 3d$_{3/2}$ and I 3d$_{5/2}$, respectively. However, the I 3d spectra and specific peak assignment for unfunctionalized UiOs and N-enriched UiOs were dramatically different. For the unfunctionalized UiO-66@I$_2$ and UiO-67@I$_2$, the I 3d peaks (632.2 eV, 619.4 eV and 633.8 eV, 619.5 eV) were mainly ascribed to I$_2$ species, owing to the weak interaction between I$_2$ and MOF host. In contrast, the I 3d spectra for iodine loaded N-enriched UiOs (UiO-66-NH$_2$@I$_2$, UiO-66-PYDC@I$_2$, UiO-67-NH$_2$@I$_2$, UiO-67-BPYDC@I$_2$) consisted of both I$_2$ (632.2 eV, 633.8 eV) and I$_3^-$ (632.2 eV, 619.4 eV) species, benefitting from the partial conversion from I$_2$ to I$_3^-$ during adsorption [50-53]. These results demonstrated that the π-conjugated electron density in N-enriched UiOs endowed a stronger charge transfer interaction between iodine and the adsorbed sites in MOF host, which thereby led to the appearance of I$_3^-$ species. Besides, it was noted that the relative proportion of 632.2 eV and 619.4 eV peak intensity was much higher for UiO-66-PYDC@I$_2$ (and UiO-67-BPYDC@I$_2$) than that for UiO-66-NH$_2$@I$_2$ (and UiO-67-NH$_2$@I$_2$). This not only suggested a dominance of I$_3^-$ species in pyridine-containing UiOs, but also indicated that the pyridine group might cause a stronger charge transfer interaction for attracting iodine molecule as compared to -NH$_2$ groups.

The N 1s spectra of iodine loaded N-enriched UiOs was also collected and depicted in **Fig. 6**g-j. Compared to as-synthesized N-enriched UiOs, the N-enriched UiOs@I$_2$ displayed higher shifted N 1s spectra, wherein the binding energy was increased by 0.1~0.8 eV. This indicated a decreased electron density around N-atom and thereby, the transferring of electron from N-atom to I-atom was experimentally evidenced for the N-enriched UiOs during iodine adsorption [54-56]. These results further confirmed that the host-guest interaction in N-enriched UiOs@I$_2$ was dominated by a strong chemical adsorption between N-functionalized groups (*e.g.*, -NH$_2$ and pyridine) and I$_2$ species.

### 3.6. DFT Calculations

DFT calculations were further applied to explore interactions between I$_2$ molecules and Zr-MOFs. Because Zr-MOFs discussed above were composed of the same Zr metal nodes linked by six different carboxylate ligands including BDC, BDC-NH$_2$, PYDC, BPDC, BPDC-NH$_2$ and BPYDC, different carboxylate ligands were extracted from Zr-MOFs and calculated to reveal their impact on I$_2$ capture capacity. The graphical representations of the optimized linkers interacting with a single I$_2$ molecule were depicted in **Fig. 7**, in which two different adsorption structures were optimized for BDC-NH$_2$, BYDC, BPDC-NH$_2$ and BPYDC linkers due to their more complexity than BDC and BPDC. As known, I₂ molecule was always inclined to be adsorbed on site with lower adsorption energy, so the situations shown in **Fig. 7**b-2, c-2, e-2 and f-2 were more reasonable than **Fig. 7**b-1, c-1, e-1 and f-1. In the following sections, we would only discuss the most reasonable adsorption situation if without special instructions.

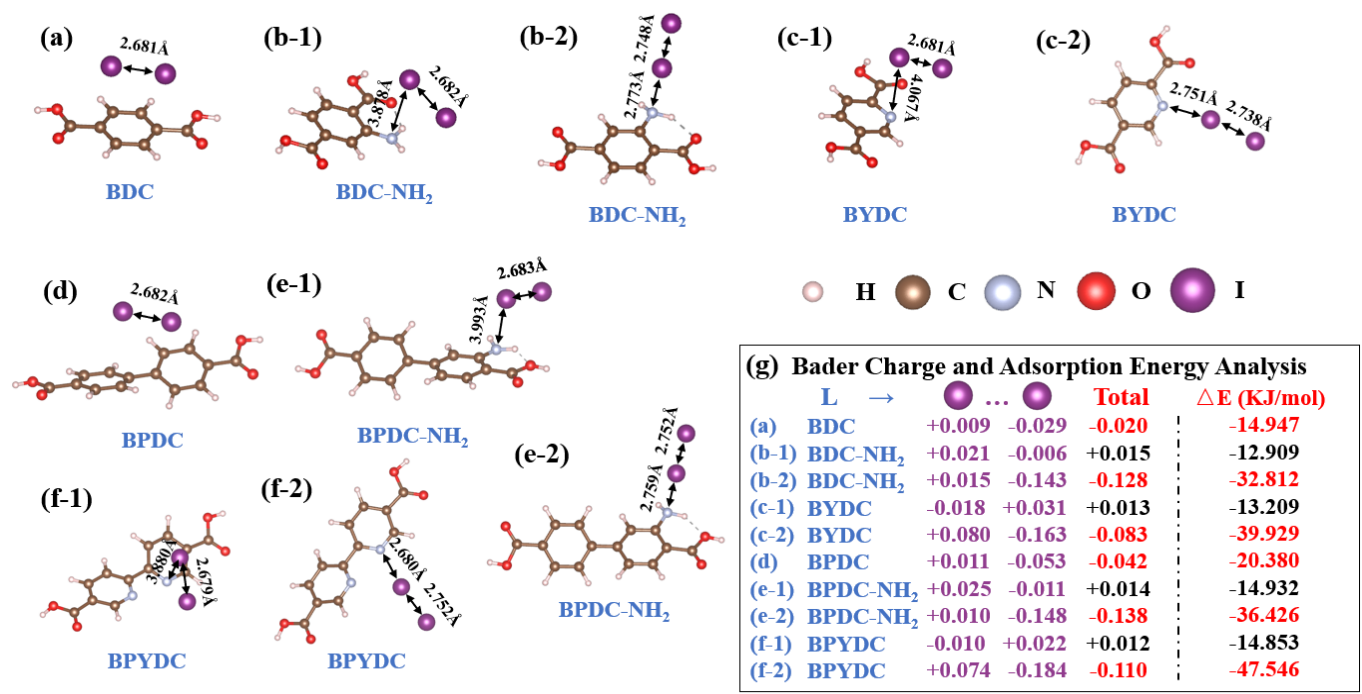

**Fig. 7.** Graphical representations of the optimized linkers interacting with a single iodine molecule for (a) BDC, (b) BDC-NH₂, (c) PYDC, (d) BPDC, (e) BPDC-NH₂, (f) BPYDC, and (g) Bader charge and adsorption energy analysis.

In **Fig. 7**a and d, the calculated bond lengths of I₂ molecules interacting with BDC and BPDC linkers were around 2.68 Å, close to the isolated I₂ molecule, which meant BDC and BPDC linkers had little affinity to iodine molecule. However, the bond lengths of iodine molecules in **Fig. 7**b-2, c-2, e-2 and f-2 were all beyond 2.73 Å, suggesting BDC-NH₂, BYDC, BPDC-NH₂ and BPYDC linkers had strong affinity to I₂ molecule. Thus, it could be concluded that the N-enriched linkers had a deeper affinity to I₂ molecules. Besides, the distances of N⋯I for BYDC and BPYDC (**Fig. 7**c-2 and f-2) were shorter than that for BDC-NH₂ and BPDC-NH₂, respectively (**Fig. 7**b-2 and e-2). This indicated a more powerful attraction towards iodine molecules from the N atoms in pyridine than that in amino. The more reasonable situations in **Fig. 7**b-2, c-2,

e-2 and f-2 than that in Fig. 7b-1, c-1, e-1 and f-1 demonstrated that the arrangement of one N-atom and $I_2$ molecule in a linear pattern was more favorable for $I_2$ adsorption than its vertical arrangement.

Bader charges were also calculated to quantitatively analyse the transferring process of charge from organic linkers to $I_2$ during adsorption (Fig. 7g). For the simple phenyl linkers of BDC and BPDC, the $I_2$ modules tended to be adsorbed on the π-conjugated electrons in benzene rings during adsorption, and a negative charge of less than 0.05 was always obtained. By contrast, for the N-containing linkers, the higher π-conjugated electron density cantered at N atoms endowed that the BDC-NH$_2$, BYDC, BPDC-NH$_2$ and BPYDC donated more negative charges (more than 0.08) to $I_2$ molecules. Compared with the simple phenyl linkers, the N-enriched linkers also showed higher adsorption energy of -32.812 kJ/mol, -39.929 kJ/mol, -36.426 kJ/mol and -47.546 kJ/mol for BDC-NH$_2$, BYDC, BPDC-NH$_2$ and BPYDC, respectively. On the whole, the N-enriched linkers exhibited a higher affinity towards $I_2$ molecules than the simple phenyl linkers in UiOs, and the pyridine-containing linkers had a stronger adsorption capacity than amino-containing linkers. In addition, the adsorption energy between the whole UiO-66 series (including UiO-66, UiO-66-NH$_2$ and UiO-66-PYDC) and iodine was calculated with two different adsorption sites considered. The difference in electronic density for UiO-66@I$_2$, UiO-66-NH$_2$@I$_2$ and UiO-66-PYDC@I$_2$ was plotted in Fig. S6, and the corresponding adsorption energy was listed in Table S6. As shown, the preferred adsorption energy for UiO-66@I$_2$, UiO-66-NH$_2$@I$_2$ and UiO-66-PYDC@I$_2$ was -60.236 kJ/mol, -97.826 kJ/mol and -109.539 kJ/mol, respectively. The above simulation results were well consistent with previous experimental results in the relatively low concentration iodine ($C_{initial}$ = 0.5 mmol/L) adsorption situation.

### 3.7. Recycling Performance

The regeneration and recycling performance of N-enriched UiOs were further researched for their utilization in the practice. For removing the adsorbed iodine species, the N-enriched UiOs@I$_2$ were firstly put into ethanol for 24 h, then centrifuged and dried at 80 °C for another 24 hours to restore the adsorption sites.

After regeneration process, the regenerated N-enriched UiOs were subjected to the multiple adsorption-desorption cycles ($C_{initial}$ = 2.5 mmol/L) for at least 4 times (**Fig. 8**). As shown, the removal efficiency of UiO-66-NH$_2$ was gradually decreased during the multiple cycles and the value was dropped from 97.1% in the 1$^{st}$ cycle to 66.7% in the 4$^{th}$ cycle (**Fig. 8**a). Although the removal efficiency variation for UiO-66-PYDC was similar to that of UiO-66-NH$_2$, the downward trend was much slower for UiO-66-PYDC and the lowest removal efficiency was still above 75% (**Fig. 8**b). Different from the above two adsorbents, it was surprising to find that UiO-67-NH$_2$ demonstrated a barely changed removal efficiency variation trend during the multiple adsorption-desorption cycles (**Fig. 8**c). The removal efficiency of UiO-67-NH$_2$ maintained at above 83%, though its first removal efficiency (87.7%) was much lower than that of UiO-66-NH$_2$ and UiO-66-PYDC. Among all the adsorbents, UiO-67-BPYDC exhibited a lowest removal efficiency during each multiple cycle (**Fig. 8**d), indicating its worst recycling performance. The PXRD patterns of spent N-enriched UiOs were plotted in Fig. S7, which confirmed that the spent adsorbents maintained their atomic structures due to good stability. The excellent stability and superior recyclability of UiO-66-PYDC and UiO-67-NH$_2$ in solutions prefigured their promising applications for I$_2$ removal from wastewaters.

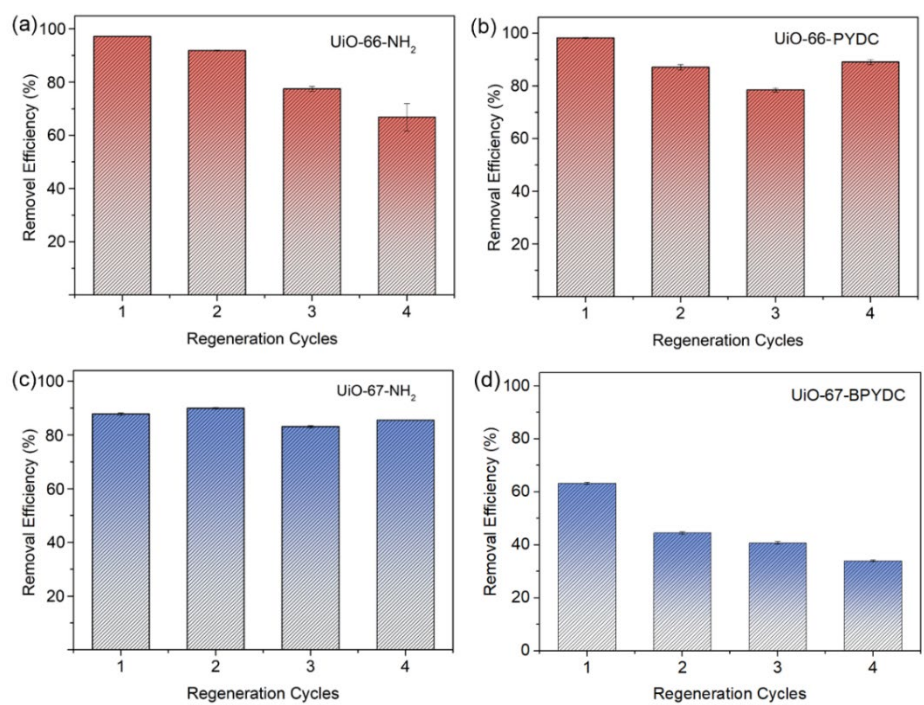

**Fig. 8.** The adsorption cycles of I$_2$/cyclohexane ($C_{initial}$ = 2.5 mmol/L) on N-enriched UiOs.

## 4. Conclusions

Two series of Zr-MOFs, including the landmark UiO-66, UiO-67 and their N-functionalized analogies (UiO-66-NH$_2$, UiO-66-PYDC, UiO-67-NH$_2$ and UiO-67-BPYDC) were synthesized and investigated for gaseous and soluble iodine adsorption under different conditions, in which the N-functionalized UiOs exhibited the faster adsorption rate and higher uptake amount. The adsorption model fitting analysis, Raman spectroscopy, XPS, and DFT calculations revealed that the N-enriched groups played a critical part in enhancing the iodine adsorption owing to the strong charge transfer effect between the N-atoms in pyridine or amino and iodine molecules. Besides, the N-functionalized UiOs (particularly, UiO-66-PYDC and UiO-67-NH$_2$) also showed good recyclability with negligible capacity reduction during recycling adsorption-desorption experiments with over four adsorption-desorption cycles. This work systematically revealed the advantage of engineering N-groups to increase the iodine affinity of adsorbents, demonstrating N-functionalized UiOs as promising candidates for the removal of radioiodine from nuclear waste.

**Acknowledgement**

This work was financially supported by the Academic Excellence Foundation of BUAA for PhD Students. The calculations in this work was carried out on TianHe-1 (A)at National Supercomputer Center in Tianjin.

## Data availability

Data will be made available on request.

**Appendix A. Supplementary material**

SEM images; PXRD patterns; Pore size distributions and textural properties of Zr-MOFs; The corresponding plots of pseudo-first-order kinetics of gaseous and soluble iodine adsorption onto Zr-MOFs; The external and intraparticle mass transfer diffusion model fitting and corresponding constants for soluble iodine adsorption on Zr-MOFs; Difference in electronic density and adsorption energy for UiO-66@$I_2$, UiO-66-NH$_2$@$I_2$ and UiO-66-PYDC@$I_2$.

Supplementary material to this article (doi:10.1016/j.mtsust.2023.100574) can be found online at https://ars.els-cdn.com/content/image/1-s2.0-S2589234723002610-mmc1.docx